\RequirePackage{fix-cm}
\documentclass[smallextended]{svjour3}       
\smartqed  
\usepackage{hyperref}
\usepackage[]{natbib}
\usepackage{url}
\makeatletter
\def\url@leostyle{%
    \def\UrlFont{\sf}}{\def\UrlFont{\small\ttfamily}}
\makeatother
\usepackage{amsmath}
\usepackage{amsfonts}

\usepackage{seqsplit}
\usepackage{tikz-cd}

\newcommand{\Tr}{\operatorname{Tr}}%
\newcommand{\ZT}[1]{\textquotedblleft#1\textquotedblright}%

\newcommand{\dif}{\mathrm{d}}%
\newcommand{\ii}{\mathrm{i}}%

\newlength{\myl}%
\newcommand{\SUM}[2]{{\setlength{\myl}{\widthof{$\displaystyle\sum_{#1}^{#2}$}*\real{0.5}-\widthof{$\displaystyle\sum$}*\real{0.5}}\sum_{#1}^{#2}\;\hspace{-\the\myl}}}
\newcommand{\INT}[3]{\settowidth{\myl}{$\displaystyle\int_{#1}^{#2}$}{\int_{#1}^{#2}\;\;\;\hspace{-\the\myl}\dif #3}\,}
\newcommand{\TINT}[3]{\settowidth{\myl}{$\int_{#1}^{#2}$}{\int_{#1}^{#2}\!\ifthenelse{\equal{#1#2}{}}{}{\;\;\;\;\hspace{-\the\myl}}\dif #3}\,}%
\newcommand{\EINT}[3]{\settowidth{\myl}{$\int_{#1}^{#2}$}{\int_{#1}^{#2}\;\;\;\,\hspace{-\the\myl}\dif #3}\,}
\usepackage{graphicx}
\usepackage{braket}
\usepackage{mathabx}
\usepackage{color}
\begin{document}

\title{Understanding probability and irreversibility in the Mori-Zwanzig projection operator formalism
}

\titlerunning{Probability and irreversibility in the Mori-Zwanzig formalism}        

\author{Michael te Vrugt}


\institute{Michael te Vrugt \at
    Institut f\"ur Theoretische Physik \\
    Center for Soft Nanoscience\\
    Philosophisches Seminar\\
    Westf\"alische Wilhelms-Universit\"at M\"unster\\
    48149 M\"unster, Germany\\
    \email{michael.tevrugt@uni-muenster.de}           
}

\date{}

\maketitle

\begin{abstract}
Explaining the emergence of stochastic irreversible macroscopic dynamics from time-reversible deterministic microscopic dynamics is one of the key problems in philosophy of physics. The Mori-Zwanzig projection operator formalism, which is one of the most important methods of modern nonequilibrium statistical mechanics, allows for a systematic derivation of irreversible transport equations from reversible microdynamics and thus provides a useful framework for understanding this issue. However, discussions of the Mori-Zwanzig formalism in philosophy of physics tend to focus on simple variants rather than on the more sophisticated ones used in modern physical research. In this work, I will close this gap by studying the problems of probability and irreversibility using the example of Grabert's time-dependent projection operator formalism. This allows to give a more solid mathematical foundation to various concepts from the philosophical literature, in particular Wallace's simple dynamical conjecture and Robertson's theory of autonomous macrodynamics. Moreover, I will explain how the Mori-Zwanzig formalism allows to resolve the tension between epistemic and ontic approaches to probability in statistical mechanics. Finally, I argue that the debate which interventionists and coarse-grainers should really be having is related not to the question why there is equilibration at all, but why it has the quantitative form it is found to have in experiments.

\end{abstract}

\section{\label{intro}Introduction}
The philosophy of statistical mechanics is a huge and rich field concerned with a variety of questions. Arguably among the most important of these are:
\begin{enumerate}
    \item What is the meaning of the probability distributions employed in statistical physics?
    \item How does macroscopic irreversibility arise from time-reversal invariant microscopic dynamics?
\end{enumerate}
Both questions are related to the observation that the microscopic equations of motion - both in classical and in quantum mechanics - are deterministic and invariant under time-reversal. Determinism appears to be in conflict with the existence of (objective) probability distributions, while time-reversal invariance appears to be in conflict with the existence of a clear arrow of time. Due to the explanatory role that probability distributions play in statistical mechanics (in particular, the thermodynamic arrow of time is often thought to be related to the initial probability distribution of the universe \citep{Wallace2011}), these problems are tied together \citep{Brown2017}.

Philosophers of physics concerned with irreversibility in statistical mechanics often focus on the history of this field, for example by analyzing the origin and meaning of Boltzmann's H-theorem. While the value of such investigations is undeniable, nonequilibrium statistical mechanics has made and is continue to make considerable progress since Boltzmann's times, and a purely historical analysis is at risk of overlooking important qualitative insights that can be gained from modern theories. In particular, nonequilibrium statistical mechanics is very successful in making quantitative predictions for the approach to equilibrium, which suggests that there is some value in what is done there despite the many \ZT{philosophical} objections against the coarse-graining methods employed there \citep{Wallace2015,Wallace2016}.

Recent attempts to close this gap on the philosophical side, in particular from the works of \citet{Wallace2015,Wallace2016} and \citet{Robertson2018}, have focused on the \textit{Mori-Zwanzig (MZ) projection operator formalism} \citep{Nakajima1958,Mori1965,Zwanzig1960}. The MZ formalism, which is one of the most important coarse-graining techniques used in modern statistical mechanics, allows for the systematic derivation of (irreversible) macroscopic transport equations based on known (reversible) microscopic dynamics. Therefore, a systematic analysis of the way in which this is done allows for an improved understanding of the origin of irreversibility in general. Philosophical discussions of the MZ formalism, however, tend to focus on very simple variants and are therefore still somewhat detached from the way in which it is used in physical research.

In this work, I will analyze the origin of probability and thermodynamic irreversibility in Grabert's time-dependent projection operator formalism \citep{Grabert1982}, which forms the basis for many applications of the MZ formalism in modern physical research. This allows for a qualitative and quantitative evaluation of various claims made concerning this formalism (and probability and irreversibility in general) in the philosophical literature. To organize the discussion, I will use a scheme proposed by \citet{teVrugt2020} in which the issue at hand is structured into five \ZT{problems of irreversibility}. 

The main conclusions of the present article are:
\begin{itemize}
\item The antagonism between epistemic and ontic approaches to probability in statistical mechanics can be traced back to the fact that the MZ formalism requires two probability distributions $\rho$ and $\bar{\rho}$. The distribution $\rho$ is the actual (quantum-mechanical) density operator of the system (as suggested by \citet{Wallace2016}), whereas the relevant density $\bar{\rho}$ is constructed on an information-theoretical basis (as suggested by \citet{Jaynes1957a}). 
\item An alternative interpretation (based on \citet{Myrvold2016}) would interpret $\rho(t)$ as the time evolute of our initial credences and $\bar{\rho}(t)$ as the convenient replacement we use for $\rho(t)$.
\item Since $\bar{\rho}$ is constructed on an information-theoretical basis in any case, we cannot avoid epistemic considerations even if we interpret $\rho$ as a physical property of the system.
\item It is shown that \citeauthor{Robertson2018}'s \citeyearpar{Robertson2018} coarse-graining criterion (revealing autonomous macro-dynamics) corresponds, in the MZ formalism, to choosing conserved quantities and broken-symmetry-variables as relevant degrees of freedom.
\item At the same time, some applications of coarse-graining in statistical mechanics are \textit{not} motivated by finding autonomous macro-dynamics, but solely by limitations of human observers. These applications are also justified.
\item It is argued that the debate between interventionists and coarse-grainers should be concerned not with whether there can be equilibration, but with whether there can be \textit{diffusive} equilibration in a closed system.
\item A quantitative analysis is provided for how the idea of forward compatibility introduced in \citet{Wallace2011} can be accomodated in the MZ formalism, and it is shown that forward compatibility requires both simple initial densities (\ZT{simple dynamical conjecture}) \textit{and} rapidly decaying memory kernels.  
\item Interpreting both $\rho(t)$ and $\bar{\rho}(t)$ as being related to the credences of observers avoids disastrous retrodictions. This is a disadvantage since there can be physical boundary conditions for which such \ZT{disastrous} retrodictions are actually correct.
\end{itemize}
This article is structured as follows: In section (\ref{problems}), I will introduce the philosophical debate concerned with probability and irreversibility in statistical mechanics. The MZ formalism is introduced in section (\ref{moriz}). In section (\ref{probmz}), I analyze the interpretation of probability in the MZ formalism. Then, I discuss in turn four \ZT{problems of irreversibility}, namely the definition of equilibrium and entropy (section (\ref{secondproblem})), coarse-graining (section (\ref{thirdproblem})), the approach to equilibrium (section (\ref{fourthproblem})), and the arrow of time (section (\ref{fifthproblem})). I conclude in section (\ref{conclusion}).




\section{\label{problems}The problem(s) of irreversibility}

\subsection{\label{probs}Probability}
A central issue in the philosophy of statistical mechanics is the understanding of probability. Since it describes systems consisting of many particles whose microscopic state is unknown, statistical mechanics operates with \textit{probability distributions}. In the classical case, the probability distribution $\rho(t)$ is typically taken to describe the probability that the system is, at a time $t$, at a certain point in phase space. In the quantum case, a system is instead described using a density operator (also known as \ZT{density matrix} or \ZT{density operator}) that in \ZT{textbook statistical mechanics} is typically introduced as
\begin{equation}
\hat{\rho}=p_i \ket{\psi}_i\bra{\psi}_i,
\end{equation}
where $p_i$ is the probability that the system is in state described by the wavefunction $\ket{\psi}_i$. Very roughly, $\hat{\rho}$ is thereby a probability distribution over wavefunctions. (From now on, I will drop the hat and refer to both the classical and the quantum distribution as \ZT{density}.)

The understanding of probability is a long-stand problem in philosophy (see \citet{Hajek2019} for a detailed review). In the philosophy of statistical mechanics (but not only there), it is common to distinguish between \textit{objectivist} and \textit{subjectivist} approaches to probability (see, e.g., \citet{Brown2017,Myrvold2012}). The debate is then concerned with whether probabilities assigned to microscopic states of a many-particle system are objective or subjective. It is then seen as a problem of \ZT{subjectivist} approaches that probabilities in statistical mechanics are often taken to have an explanatory role in, e.g., the approach to thermodynamic equilibrium, which is an objective feature of the world \citep{Albert1994,Albert1994b}. On the other hand, one might wonder where an \ZT{objective probability distribution} might come from given that the microscopic dynamics is deterministic \citep{Brown2017}.


A part of the problem is that it is not entirely clear what one could mean by an \ZT{objective probability distribution} in the context of statistical mechanics. \citet{Brown2017} argues that the typical definition of \ZT{objective probability} in terms of relative frequencies is problematic in many ways, but that there also appears to be no convincing alternative. \citet{vonKutschera1969} comes to a similar conclusion regarding the frequentist approach and argues that a proper interpretation of \ZT{objective probability} in the natural sciences should include a subjective element. \citet{Myrvold2012} argues that the dichotomy between objective and subjective probabilities does no justice to statistical mechanics and therefore proposes to introduce \ZT{almost objective probabilities} (see below) based on human credences, but admits that these by itself will not directly explain any thermodynamic behavior (which is what probability distributions in statistical mechanics are often used for). And \citet[p. 12]{Wallace2016} comes to the conclusion that (in classical statistical mechanics) objective probability is \ZT{a mysterious concept}. It thus appears as if \ZT{objective} and \ZT{subjective} probability are maybe not the concepts we should distinguish between.

A more precise terminology would be to distinguish (as done, e.g., by \citet{Frigg2008}) between \textit{epistemic probabilities} and \textit{ontic probabilities}. Epistemic probabilities represent degrees of belief (these can be \textit{objective}, i.e., the same for every rational observer with the same evidence, or \textit{subjective}, i.e., not fully determined by the evidence), where as ontic probabilities represent aspects of the physical world (representing frequencies, propensities, or being part of a Humean best system). When we argue that a probability should be objective in order to be able to play an explanatory role in physics, what we really mean is that it should be a \textit{physical} (ontic) probability since only physical circumstances can explain physical effects\footnote{I leave aside here the question whether there can be non-physical causes of physical effects (such as minds or gods).} A probability assignment based on degrees of belief might be objective in the sense that (almost) every rational agent has the same degree of belief in a certain situation, but these degrees of belief still cannot explain any process in the real world. 

I now present in more detail three ways in which the probabilities used in statistical mechanics can be understood\footnote{This is by no means an exhaustive list of the options suggested in the literature. There is, for example, an interesting approach based on incorporating probabilities in a Humean best system \citep{FriggH2015,Frigg2016}. Here, I have chosen three options that fit particularly well to the mathematical formalism discussed in section (\ref{mz}).}:
\begin{enumerate}
    \item The probabilities in statistical mechanics arise from quantum-mechanical probabilities \citep{Albert1994,Albert1994b,Wallace2016}.
	\item The probabilities in statistical mechanics represent our knowledge about the system \citep{Jaynes1957a,Jaynes1957b}.
    \item The probabilities in statistical mechanics are \ZT{almost objective probabilities} \citep{Myrvold2012,Myrvold2016}.
\end{enumerate}









The first option comes in different forms. I will discuss two of them here. First, David \citet{Albert1994,Albert1994b,Albert2000} has suggested that the spontaneous collapses of the wavefunction postulated by Ghirardi-Rimini-Weber (GRW) theory \citep{GhirardiRW1986} could allow to explain thermodynamic irreversibility. In the GRW theory, it is assumed that the wavefunction of a particle is, at a certain rate, multiplied by a Gaussian. This leads to a localization and (effectively) to a collapse. Since GRW theory involves objective stochasticity, one would then have a straightforward explanation for the existence of objective probabilities in statistical mechanics. However, recent computer experiments by \citet{teVrugtTW2021} indicate that Albert's suggestion is not successful as an explanation of thermodynamic irreversibility even if GRW theory is true. \citet{Albert1994,Albert1994b} argues that the GRW collapses will bring a system starting in an \ZT{abnormal} initial state\footnote{An abnormal initial state is a state that, if evolved forward in time using Hamiltonian dynamics, leads to anti-thermodynamic behavior.} into a state that leads to a normal thermodynamic time evolution. However, \citet{teVrugtTW2021} have found no such effect in their simulations.

Second, David \citet{Wallace2016} has argued that the probabilities of statistical mechanics arise from \ZT{standard} quantum mechanics without spontaneous collapses. His starting point is the observation that the interpretation of the density operator as a \ZT{probability distribution over wavefunctions} is not generally possible. If we consider a system consisting of two particles (or, more generally, subsystems) A and B, then A and B will typically be entangled, i.e., the wavefunction describing the state of the joint system cannot be written as a product of a wavefunction for A and a wavefunction for B. In fact, there is no wavefunction that can describe all possible measurements on A, and the correct description of the state of A is a density operator obtained by taking the trace over the degrees of freedom of B. This density operator, however, cannot be interpreted as a probability distribution over possible states of A. Consequently, the use of density operators in quantum mechanics is required because of entanglement regardless of any considerations about probability, such that \ZT{there is nothing formally novel about their introduction in statistical mechanics} \citep[p. 18]{Wallace2016}. Moreover, a probability distribution over mixed states is mathematically indistinguishable from an individual mixed state. Thus, given that essentially all systems of interest to statistical mechanics are entangled with their environment, we can interpret the mixed states used in (quantum) statistical mechanics as states of \textit{individual systems} rather than as probability distributions over possible pure states.

Wallace's approach is far more promising than Albert's in the understanding of the quantum origins of statistical mechanics, in particular because it is in line with recent progresses in quantum thermodynamics. I discuss the latter following the review by \citet{VinjanampathyA2016}. As they point out, statistical mechanics describes the universe using a mixed state where every state compatible with the total energy $E$ of the universe has the same probability, which is in contradiction with the requirement that the quantum state of the universe is pure\footnote{Note that philosophers of physics (such as \citet{Chen2020}) sometimes deny that the universe has to be in a pure state.}. \citet{PopescuSW2006} showed that the density operator of a subsystem of the universe, obtained by tracing the actual density operator of the universe over all other degrees of freedom, is very close in trace distance to the state obtained by tracing out the maximally mixed state. In other words, a sufficiently small subsystem cannot distinguish between a mixed and a pure universe. This theorem, referred to as \ZT{general canonical principle}, is supposed to replace the typical postulate of equal a priori probabilities.

I now turn to the second option, which I explain following \citet{Frigg2008}. According to \citet{Jaynes1957a,Jaynes1957b}, the probability distributions of statistical mechanics represent our knowledge about a system. By the principle of indifference, equal probabilities are to be assigned to all microstates that are compatible with the macroscopic evidence. Suppose that a random variable $x$ is continuous and can take values in an interval $[a,b]$. Then, the probability distribution $p(x)$ should be chosen in such a way that it maximizes the Shannon entropy
\begin{equation}
S_\mathrm{S}=-\INT{a}{b}{x}p(x)\ln(p(x))
\end{equation}
subject to macroscopic constraints of the form
\begin{equation}
\braket{f}= \INT{a}{b}{x}f(x)p(x) =c,
\end{equation}
which express that (according to our macroscopic evidence) the mean value of an observable $f$ is equal to $c$. Notably, although this approach is commonly denoted \ZT{subjectivist}, the probabilities in Jaynes' theory are determined by the available data and do therefore not (solely) represent the personal opinions of individual observers. Thus, \ZT{epistemic} is the more appropriate terminology \citep{Frigg2008}.

The third approach, defended by \citet{Myrvold2016,Myrvold2012}, is based on the method of arbitrary functions. Here, the idea is that the time evolutes of an (almost) arbitrary initial probability distribution will give the same results for the probabilities of certain (macroscopically feasible) measurements. These probabilities then are \ZT{almost objective}. As an example, suppose that a gas is initially (at time $s$) confinend to the left half of a box and then allowed to expand, and that an agent Alice has some credence regarding the state of the gas (represented by a certain probability distribution $\rho_A(s)$). Let $\rho_A(t)$ be the result of evolving $\rho_A(s)$ forward in time using the Liouville equation to a time $t$ (sufficiently long after $s$ to allow for equilibration). Typically, $\rho_A(t)$ will be extremely complicated. The same holds for $\rho_B(t)$, which is the result of evolving forward in time the initial credences $\rho_B(s)$ of an agent Bob who believes that the gas was initially confined to the right of the box. However, $\rho_A(t)$, $\rho_B(t)$, and the equilibrium distribution $\rho_{\mathrm{eq}}$ will give the same probabilities for all macroscopic measurements. In fact, these equilibrium probabilities arise from almost every initial credence function (all except for those that would require very detailed knowledge about then microscopic state), making these probabilities almost objective \citep{Myrvold2012}.

\subsection{\label{coarse}Coarse-graining}
Suppose now that we have found a probability distribution $\rho$ on phase space (in the classical case) or a density operator $\rho$ (in the quantum case) that describes our system. One can then define the \textit{Gibbs entropy}\footnote{Gibbs, being a frequentist, thought of the probability $\rho$ as measuring the fraction of systems in an ensemble (hypothetical set of infinitely many copies of a system) that are in a particular state \citep[p. 585]{Myrvold2016}. Nowadays, one typically distinguishes between Gibbsian and Boltzmannian statistical mechanics, with the former being based on ensembles. This issue is not essential for this article, see \citet{Myrvold2016,Frigg2008} for a discussion of this distinction and its relation to the problem of probabilities.} as
\begin{equation}
S = - k_B \Tr(\rho\ln\rho),   
\label{finegrained}
\end{equation}
where $k_B$ is the Boltzmann constant and the trace $\Tr$ denotes a phase-space integral in classical and a quantum-mechanical trace in quantum mechanics. In the quantum case, the entropy defined by Eq.\ (\ref{finegrained}) is the \textit{von Neumann entropy}. It is common in physics to identify the von Neumann entropy with the thermodynamic entropy, whereas this is slightly controversial in philosophy (see \citet{HemmoS2006,Shenker1999} for arguments against and \citet{Chua2021,Henderson2003} for arguments in favor of this view). An important property of both the classical and the quantum Gibbs entropy is that it is constant in a system governed by Hamiltonian mechanics (which is invariant under time reversal).

This is problematic since in macroscopic thermodynamics the entropy is \textit{not} constant. It increases and reaches its maximum in the equilibrium state. Thus, a central challenge of Gibbsian statistical mechanics is to make the entropy (\ref{finegrained}) change \citep{Frigg2008}. A common way to do this is to replace $\rho$ by an \ZT{averaged} density $\bar{\rho}$ that is typically referred to as the \textit{coarse-grained density}. (In contrast, $\rho$ is then called the \textit{fine-grained density}.) If we replace $\rho$ by $\bar{\rho}$ in Eq.\ (\ref{finegrained}), we get the \textit{coarse-grained entropy} (as opposed to the \textit{fine-grained entropy} defined in terms of $\rho$). Unlike the fine-grained one, the coarse-grained entropy can increase. A typical justification for the replacement $\rho\to\bar{\rho}$ is that this replacement corresponds to ignoring microscopic details that we cannot measure anyway; a common objection is then that any irreversibility that results from this coarse-graining is an artefact that is therefore illusory and/or anthropocentric \citep{Robertson2018}.

The coarse-graining procedure can be most easily understood in the classical case, where it is commonly discussed using an analogy by Gibbs (see \citet[pp. 549-551]{Robertson2018}). The density $\rho$ evolves like an incompressible fluid in phase space, much like a drop of ink put into water. Since the ink is incompressible, its volume remains constant. However, as it fibrillates through the water and thereby makes the whole water look blue, it appears as if the ink's volume has increased, and it has if we measure it only in an averaged way.

Although the ink analogy provides a very good illustration, one important aspect is easily overlooked: While coarse-graining ensures that the entropy \textit{can} increase, it does by no means imply that it \textit{has} to. To stay in the ink picture: The ink may spread out (and this is what it is found to do in practice), but it is also logically conceivable that it stays where it is or even shrinks back to the origin, which would correspond to a decrease of entropy. Moreover, nothing guarantees a priori that the ink spreads out monotonically rather than, for example, oscillatory. Consequently, while coarse-graining may be a choice that humans make due to a finite measurement resolution, the fact that one observes, for a given coarse-graining, an increase of the entropy is still a substantial physical statement. 

To understand this situation in more detail, we should take into account that the aforementioned \ZT{problem of irreversibility} actually consists of a variety of sub-problems. Following \citet{teVrugt2020}, I distinguish between five \ZT{problems of irreversibility}:
\begin{itemize}
    \item \textbf{Q1:} What is the location of irreversibility within thermodynamics?
    \item \textbf{Q2:} What is the definition of \ZT{equilibrium} and \ZT{entropy}?
    \item \textbf{Q3:} What is the justification of coarse-graining?
    \item \textbf{Q4:} Why do systems that are initially in a nonequilibrium state approach equilibrium?
    \item \textbf{Q5:} Why do system approach equilibrium in the future, but not in the past?
\end{itemize}

Q1 is concerned entirely with the macroscopic theory of phenomenological thermodynamics, and asks which of its axioms actually makes the theory irreversible (see \citet{BrownU2001,Luczak2017,Robertson2021}). Q2 asks how we should define the entropy, in particular whether we should define it in terms of $\rho$ or in terms of $\bar{\rho}$ (which, if we assume the equilibrium state to be the one with the maximal entropy, also implies different definitions of \ZT{equilibrium}). Q3 then asks why it is justified to replace $\rho$ by $\bar{\rho}$. Since this replacement not yet implies an increase of entropy, Q4 then asks why the entropy increases (and not, for example, decreases or remains constant). And since this explanation will, due to the time-reversal symmetry of the underlying microdynamics, often also be applicable to the past, Q5 finally is concerned with why entropy increase only takes place in one direction of time. A common answer to Q5 is the \ZT{past hypothesis}, in which the thermodynamic asymmetry is explained via an assumption about the initial state of the universe (typically the assumption that the entropy of the early universe was very low) \citep{Albert2000}. (See \citet{Frisch2005,Wallace2011,Brown2017,Farr2021} for a further discussion of the past hypothesis.)

\section{\label{moriz}Mori-Zwanzig formalism}

\subsection{\label{mzp}The Mori-Zwanzig formalism in philosophy}

Having discussed the general theory of coarse-graining, we now come to one of the most important coarse-graining methods used in modern physics, namely the \textit{Mori-Zwanzig (MZ) projection operator formalism}, developed by \citet{Mori1965}, \citet{Zwanzig1960}, and \citet{Nakajima1958}. It has a large number of applications in modern physics, including (but not limited to) active matter \citep{HanFSVIdPV2020}, dynamical density functional theory \citep{EspanolL2009,teVrugtLW2020}, general relativity \citep{teVrugtHW2021}, glasses \citep{Das2004}, high-energy physics \citep{HuangKKR2011}, and solid-state theory \citep{Fulde1995}. Introductions to the formalism can be found in \citet{Grabert1982,teVrugtW2019b,RauM1996,KlipensteinTJSvdV2021,Schilling2021}. Essentially, the MZ formalism allows to describe the dynamics of a many-particle system in terms of the closed subdynamics of an arbitrary set of \ZT{relevant variables} $\{A_j\}$. The central idea here is that all variables that can be used to describe the system form a Hilbert space, and the relevant variables form a subspace. (This Hilbert space of observables, which is a convenient mathematical construction, is not to be confused with the Hilbert space of quantum states.) One can now introduce a scalar product and, based on this, a projection operator that allows to project the full dynamics onto the subspace of the relevant variables. The irrelevant part of the dynamics then enters the dynamics via memory and noise terms. As a result, one gets a closed and exact transport equation for the relevant variables. If one approximates the memory term by a memoryless contribution, one gets irreversible dynamics. Consequently, the MZ formalism provides a highly useful tool for studying the microscopic origins of thermodynamic irreversibility \citep{teVrugtW2019b}.

This usefulness has not gone unnoticed in the foundations of physics. In his book on the arrow of time in physics, \citet{Zeh1989} has devoted a chapter to the MZ formalism, analyzing in detail how it leads from reversible to irreversible dynamics and which assumptions are involved there. A  shorter and less technical discussion was provided by \citet{Sklar1995}. \citet{RauM1996} have provided a detailed review of how the MZ formalism allows to study the emergence of irreversibility. Later, \citet{Wallace2015,Wallace2016} has discussed the MZ formalism as a paradigmatic case of a quantitative method in nonequilibrium statistical mechanics. Finally, \citet{Robertson2018} has used this formalism (which she refers to as \ZT{Zwanzig-Zeh-Wallace (ZZW) framework}\footnote{This terminology will not be used here for two reasons. First, the name \ZT{Mori-Zwanzig formalism} is \textit{way} more common, in particular in the physics literature. Second, the name \ZT{ZZW} framework gives credit not only to authors who developed the formalism (Zwanzig), but also to those who \ZT{only} gave an analysis of the formalism in the context of irreversibility (Zeh and Wallace).}) as a basis for the position that coarse-graining in statistical mechanics should aim at revealing autonomous macroscopic dynamics.

\citet{Wallace2011} develops the following mathematical understanding of coarse-grained dynamics: In general, a microscopic density $\rho$ can be evolved forwards in time using the microscopic dynamics $U$. For any coarse-graining procedure $C$, one can define the $C+$ dynamics as follow: Apply coarse-graining to the microscopic distribution, evolve it for a short time $\Delta t$ using the microdynamics, coarse-grain again, evolve for $\Delta t$ again etc. A distribution $\rho$ is said to be forward compatible with $C$ if evolving it using $U$ and then coarse-graining at the end gives the same result as evolving it using the $C+$ dynamics. Hence, an initial density $\rho(s)$ is forward compatible if the diagram
\[
\begin{tikzcd}
\rho(s)\ar[r, "U"] \ar[d,"P^\dagger"] & \rho(t)\ar[d,"P^\dagger"]\\
\bar{\rho}(s)\ar[r,"C+"] & \bar{\rho}(t)
\end{tikzcd}
\]
commutes. (Diagram adapted from \citet[p. 557]{Robertson2018}.) \citet{Wallace2011} then introduces the \textit{simple dynamical conjecture}, which states that any distribution that is \ZT{simple} is forward compatible\footnote{The original statement \cite[p. 19]{Wallace2011} uses \ZT{forward predictable} (which is a slightly stronger requirement defined in \citet{Wallace2011}) instead of \ZT{forward compatible}. Here, I follow \citet{Robertson2018}, who frames her discussion exclusively in terms of forward compatiblity. Since forward predictability implies forward compatibility (and since forward compatibily implies forward predictability for macrodeterministic systems) \citep[p. 16]{Wallace2011}, the simple dynamical conjecture as stated here follows from the original formulation and is equivalent with it for most cases of practical relevance. Note that the mathematical analysis in section (\ref{fifthproblem}) is based on the definition used here.} with $C$. He does not give a precise definition of simplicity, but suggests that a distribution that can be specified in a closed form as a uniform distribution over certain macroproperties is simple whereas one that can be specified only by time-evolving another distribution\footnote{An example for such a state would be the one shown in Fig. 1b of \citet{teVrugtTW2021}.} is not \citep[p. 19]{Wallace2011}. Based on the simple dynamical conjecture, Wallace then introduces the \textit{simple past hypothesis} which, in the quantum-mechanical form, assumes that the initial quantum state of the universe is simple. This then explains the physical arrow of time. (Note that this initial quantum state is not necessarily pure.)

A more specific discussion of the MZ formalism can be found in \citet{Wallace2015,Wallace2016}. Here, Wallace introduces it as a prototypical example of coarse-graining in nonequilibrium statistical mechanics and presents the standard derivation of the master equation (which is shown here in our notation). The microscopic density $\rho$ obeys
\begin{equation}
\dot{\rho}(t)= -\ii L \rho(t), 
\label{liouville}
\end{equation}
where $L$ is the Liouvillian (defined as $L = \frac{1}{\hbar}[H,\cdot]$ with the reduced Planck constant $\hbar$) and the dot denotes a time derivative. One defines a projection operator $P^\dagger$ and an orthogonal projection operator $Q^\dagger$ with the property $P^\dagger\rho = \bar{\rho}$, where $\bar{\rho}$ is the relevant part of the density. (We use, following the notation in \citet{Grabert1982}, the dagger to distinguish the projection operator $P^\dagger$, which acts on density operators, from the projection operator $P$ introduced later, which acts on observables. The operators $P$ and $P^\dagger$ are simply each others adjoint, and we can calculate $P^\dagger$ explicitly once we know $P$ \citep[p. 16]{Grabert1982}.) This allows, defining\footnote{Most philosophers write $\rho_{\mathrm{rel}}$ and $\rho_{\mathrm{ir}}$ rather than $\bar{\rho}$ and $\delta\rho$. The former notation, however, is more common in physics and is in particular used by \citet{Grabert1982} whose method this article is based on.} $\delta\rho=\rho-\bar{\rho} = Q^\dagger\rho$, to derive the following exact transport equation for $\rho$ (see \citet[p. 292]{Wallace2015} and \citet[p. 62]{Zeh1989}):
\begin{equation}                          
\dot{\bar{\rho}}(t)=-P^\dagger\ii L \bar{\rho}(t) + \INT{0}{t}{u}P^\dagger\ii L e^{-Q^\dagger\ii L u}Q^\dagger\ii L \bar{\rho}(t-u) - P^\dagger\ii L e^{-Q^\dagger\ii L t}\delta\rho(0).
\end{equation}
Setting $\delta\rho(0)=0$ and assuming that the memory kernel vanishes rapidly (Markovian approximation) gives the time-irreversible approximate transport equation (master equation) \citep[p. 292]{Wallace2015}
\begin{equation}
\dot{\bar{\rho}}(t)=-P^\dagger\ii L \bar{\rho}(t) + \bigg(\INT{0}{\infty}{u}P^\dagger\ii L e^{-Q^\dagger\ii L u}Q^\dagger\ii L\bigg) \bar{\rho}(t).
\label{higherlevel}
\end{equation}
In particular, \citet[p. 292]{Wallace2015} emphasizes the importance of the assumption $\delta\rho(0)=0$, which is a probabilistic assumption about the initial state of the system. \citet[p. 61]{Zeh1989} has compared this way of eliminating the irrelevant degrees of freedom to the way in which one eliminates the \ZT{advanced} solutions in the theory of electromagnetic waves. (See \citet{Frisch2005b,Frisch2006} for a discussion of the relation between the thermodynamic and the electromagnetic arrow of time.) Note that the term $P^\dagger\ii L \bar{\rho}$ often vanishes \citep[p. 62]{Zeh1989}.

Based on these considerations, \citet{Robertson2018} has developed a theory of the justification of coarse-graining in statistical mechanics. She argues \citep[p. 556]{Robertson2018} that this procedure can be justified in three ways - interventionism (the environment implements the projection $P^\dagger$), asymmetric microscopic laws (dynamically ensuring $\rho \to \bar{\rho}$) and special initial conditions (ensuring that the coarse-grained dynamics gives the correct results for the relevant part of the density). She focuses on the third strategy (\ZT{special conditions account}). Typically, Robertson argues, coarse-graining is justified based on measurement imprecisisions (we do not know the exact microstate of a system and therefore have to use an averaged, i.e., coarse-grained, description). Based on this, it is frequently objected that the irreversible laws obtained by coarse-graining are anthropocentric and/or illusory. However, Robertson continues, we do in fact not coarse-grain because of measurement imprecisions but because we want to reveal autonomous higher-level macrodynamics. This is what the MZ formalism does, and the projection operator $P^\dagger$ has to be constructed in such a way that it leads to such autonomous dynamics. Consequently, the irreversibility of the higher-level transport equation (\ref{higherlevel}) is not illusory, but (weakly) emergent.

\subsection{\label{mz}Grabert's projection operator formalism}
Wallace and Robertson, while in principle acknowledging the broad applicability of the MZ formalism, have in practice only considered one rather simple variant, namely the derivation of master equations using time-independent projection operators. Hence, one is not explicitly concerned here with individual observables (foundational discussions usually don't even mention the Hilbert-space understanding of the MZ formalism). While the master equation approach is quite general \citep{Grabert1982}, it is also highly abstract, which has the consequence that studying only this variant leads one to overlooking important issues. Current research on this topic in physics instead focuses on studying the dynamics of individual observables using time-dependent projection operators. These variants of the formalism have been pioneered by \citet{Robertson1966}, \citet{KawasakiG1973} and \citet{Grabert1978}. More recently, extensions have been derived by \citet{MeyerVS2017,MeyerVS2019} and \citet{teVrugtW2019}. Here, I will explain the time-dependent projection operator formalism as it is described in the textbook by \citet{Grabert1982}, which forms the basis of most of the work that is done today.

The microscopic description of a many-particle system is given by its density operator $\rho$. Since it is not known exactly, it has to be approximated. For this purpose, one introduces a \textit{relevant density} $\bar{\rho}$. The relevant density is often assumed to have the form \citep[p. 482]{Grabert1978}
\begin{equation}
\bar{\rho}(t)=\frac{1}{Z(t)}e^{-a_i^\natural(t)A_i}
\label{relevantdensity}
\end{equation}
where the partition function $Z(t)$ is a normalization constant (ensuring $\Tr(\bar{\rho}(t))=1$) and the thermodynamic conjugates $a^\natural(t)$ (this notation is adapted from \citet{WittkowskiLB2012,WittkowskiLB2013}) ensure that the macroequivalence condition $a_i(t)=\Tr(\bar{\rho}(t)A_i)$ holds. (Summation over indices appearing twice is assumed throughout this article.) Here,
\begin{equation}
a_i(t)=\Tr(\rho(t)A_i)
\label{average}
\end{equation}
is the average of the observable $A_i$.
The form \eqref{relevantdensity} can be justified from an information-theoretical point of view as it expresses maximal noncommittance regarding microscopic details, i.e., it assigns all microscopic configurations that are compatible with the set of macroscopic values $\{a_i(t)\}$. 

One can then define the time-dependent projection operator $P$ acting on an arbitrary observable $X$ as \citep[p. 487]{Grabert1978}
\begin{equation}
P(t)X= \Tr(\bar{\rho}(t)X) + (A_i - a_i(t))\Tr\bigg(\frac{\partial\bar{\rho}(t)}{\partial a_j(t)}X\bigg). 
\label{projectionoperator}
\end{equation}
Equation (\ref{projectionoperator}) allows, by taking the adjoint, to get an explicit expression for the projection operator $P^\dagger$ acting on a density operator $\rho$, namely \citep[p. 16]{Grabert1982}
\begin{equation}
P^\dagger(t)\rho = \bar{\rho}(t)\Tr(\rho)+ (\Tr(A_i\rho)-a_i(t)\Tr(\rho)) \frac{\partial\bar{\rho}(t)}{\partial a_j(t)},
\end{equation}
implying
\begin{equation}
P^\dagger(t)\rho(t)=\bar{\rho}(t).
\label{property}
\end{equation}
From the Liouville equation (\ref{liouville}), which holds in the Schr\"odinger picture, one can derive the Heisenberg picture equation of motion
\begin{equation}
\dot{A}_i = \ii L A_i,    
\end{equation}
which (for a time-independent Liouvillian $L$) has the solution
\begin{equation}
\dot{A}_i(t)=e^{\ii L t}A_i,    
\label{formalsolution}
\end{equation}
where we write $A_i(0)=A_i$ (i.e., we assume Schr\"odinger and Heisenberg picture to coincide at time $t=0$ \citep{BalianV1985}). We can now insert the operator identity \citep[p. 16]{Grabert1982}
\begin{equation}
e^{\ii L t} = e^{\ii L t}P(t) + \INT{s}{t}{u}e^{\ii L u}P(u)(\ii L - \dot{P}(u))Q(u)G(u,t) + e^{\ii L s}Q(s)G(s,t)
\end{equation}
with the orthogonal dynamics propagator
\begin{equation}
G(s,t)= \exp_R\bigg(\INT{s}{t}{u}\ii LQ(u)\bigg)
\end{equation}
and the right-time-ordered exponential $\exp_R$ (see \citet{teVrugtW2019}) into Eq.\ (\ref{formalsolution}) and average the result over $\rho(0)$\footnote{Since this derivation works in the Heisenberg picture, the density operator is constant and we can use its initial form $\rho(0)$ to get the correct average at all times.}. As a result, we find an \textit{exact} dynamic equation for the mean values $a_i$, which reads 
\citep[p. 19]{Grabert1982}
\begin{equation}
\dot{a}_i(t)= v_i(t) + \INT{s}{t}{u} K_i(t,u) + f_i(t,s)
\label{gener}
\end{equation}
with the organized drift
\begin{equation}
v_i(t)=\Tr(\bar{\rho}(t)\dot{A}_i),  
\label{organizeddrift}
\end{equation}
the memory function
\begin{equation}
K_i(t,u)=\Tr(\bar{\rho}(u)\ii L Q(u)G(u,t)\dot{A}_i),
\label{ki}
\end{equation}
and the mean random force
\begin{equation}
f_i(t,s)=\Tr(\delta\rho(s)G(s,t)\dot{A}_i).
\label{meanrandomforce}
\end{equation}
Up to now, we have not used the fact that $\bar{\rho}$ has the form (\ref{relevantdensity}). If we now choose it to have this form, we can use the relation
\citep[p. 483]{Grabert1978}
\begin{equation}
-\ii L \bar{\rho}(t)=a^\natural_j(t)\INT{0}{1}{\alpha}e^{-\alpha a^\natural_k A_k}\dot{A}_je^{\alpha a^\natural_k A_k}\bar{\rho}(t),
\label{fuenf}
\end{equation}
to get \citep[p. 33]{Grabert1982}
\begin{equation}
K_i(t,u) = R_{ij}(t,u)a_j^\natural(u)
\label{kr}
\end{equation}
with the retardation matrix
\begin{equation}
R_{ij}(t,u)=\INT{0}{1}{\alpha}\Tr(\bar{\rho}(u)e^{\alpha a^\natural_k(u)A_k}(Q(u)G(u,t)\dot{A}_i)e^{-\alpha a^\natural_k(u)A_k} \dot{A}_j).    
\label{retardationmatrix}
\end{equation}
Inserting Eq.\ (\ref{kr}) into Eq.\ (\ref{gener}) gives (setting $s=0$)
\begin{equation}
\dot{a}_i(t)= v_i(t) + \INT{s}{t}{u} R_{ij}(t,u)a_j^\natural(u) + f_i(t,s).
\label{exact}
\end{equation}
We now make two important assumptions:
\begin{enumerate}
    \item Markovian approximation: The relevant variables evolve slowly compared to the microscopic degrees of freedom. This implies that the memory kernel in Eq.\ (\ref{exact}) falls off on a very short timescale, and that the thermodynamic conjugates $a_i^\natural(t)$ are approximately constant on this timescale. 
    \item The density operator at the initial time $t=s$ is of the relevant form, i.e., $\delta\rho(s)=0$. This allows to set $f_i=0$ (as can be seen from Eq.\ (\ref{meanrandomforce})).
\end{enumerate}
This allows to replace Eq.\ (\ref{exact}) by the approximate transport equation
\begin{equation}
\dot{a}_i(t)= v_i(t) + D_{ij}(t)\frac{\partial S}{\partial a_j(t)}
\label{markov}
\end{equation}
with the diffusion tensor
\begin{equation}
D_{ij}(t)=\frac{1}{k_B}\INT{0}{\infty}{u}\INT{0}{1}{\alpha}\Tr(\bar{\rho}(t)e^{\alpha a^\natural_k(t) A_k}(e^{\ii L u}Q(t)\dot{A}_i)e^{-\alpha a^\natural_k(t) A_k}\dot{A}_j).
\end{equation}
Formally, the Markovian approximation corresopnds to disregarding terms of third or higher order in $\dot{A}_i$ (which is what allows us to replace $a_j^\natural(u)$ by $a_j^\natural(t)$) \citep[p. 39]{Grabert1982}. We have used that
\begin{equation}
a^\natural_j =  \frac{1}{k_B}\frac{\partial S}{\partial a_j(t)}   
\end{equation}
with the (coarse-grained) entropy \citep[p. 483]{Grabert1978}
\begin{equation}
S = - k_B\Tr(\bar{\rho}\ln\bar\rho)= k_B\ln Z + k_B a^\natural_j a_j.
\label{entropy}
\end{equation}
One can show that
\begin{enumerate}
    \item the organized drift term $v_i$ does not contribute to the rate of change of the entropy \citep[pp. 483-484]{Grabert1978}.
    \item the tensor $D_{ij}$ is, due to the Wiener-Khintchine theorem, positive definite \citep{EspanolL2009,AneroET2013}.
\end{enumerate}
This implies that
\begin{equation}
\dot{S} = k_B^2 D_{ij} a^\natural_i a^\natural_j  \geq 0,  
\label{htheorem} 
\end{equation}
which shows that the approximate transport equation (\ref{markov}) is irreversible \citep{AneroET2013}.

\section{\label{probmz}Probability in the Mori-Zwanzig formalism}
I will now explain what the MZ formalism can contribute to solving the problems of irreversibility and probability in statistical mechanics. On the one hand, I will thereby provide a conceptual understanding of the mathematical formalism outlined in section (\ref{mz}). Moreover, I will use the equations from section (\ref{mz}) in order to give the general considerations from Wallace and Robertson a quantitative underpinning.

First, I will discuss how, within the MZ formalism, we can address the problem of understanding probability in statistical mechanics as introduced in section (\ref{probs}). The key point to take into account here is that there are \textit{two} densities, not one. First, there is the microscopic density $\rho$. It describes (at least according to \ZT{textbook understanding}) the actual state of the system. Second, there is the relevant density $\bar{\rho}$ (typically given by Eq.\ (\ref{relevantdensity})). It describes our knowledge about the state of the system. Although the existence of these two different distributions is never questioned in the physics literature, it is actually extremely surprising from a classical point of view. Both $\rho$ and $\bar{\rho}$ are probability distributions. If it is the point of probability distributions in statistical mechanics to express ignorance of the system, then what does \ZT{correct microscopic probability distribution} even mean? And if this is not the point of probabilities in statistical mechanics, then what is?

In section (\ref{mz}), I have introduced three interpretations of probability in statistical mechanics. Let us start with the first one by taking $\rho$ to be simply the actual density operator of the system. There are two important objections against this view. First, one could ask whether using the density operator really allows us to get probabilities. This has been questioned by \citet[p. 38]{Brown2017}, who argued that, within the Everettian interpretation of quantum mechanics that Wallace defends, \ZT{a density operator (...) is no more intrinsically a carrier of probability than is the Liouville measure on the classical phase space.} Probabilities, he argues, arise - in an Everettian framework - only when a rational agent bets on measurement outcomes\footnote{It is common in the Everett interpretation to assume that quantum-mechanical probabilities are related to the betting behavior of rational agents \citep{Wallace2012}.}. This implies that (for an Everettian) \ZT{quantum probabilities make no appearance at the start of the world, but are forced on us at the later times at which observations are made} \citep[p. 38]{Brown2017}, and thus seems to suggest that quantum probabilities do not give us statistical probabilities at the start of the world. While Brown's observation is correct, it does not pose any problem for Wallace's approach (at least not when it is applied to the MZ formalism). The only reason we require an interpretation of $\rho$ as (something like) a probability distribution is that we want to interpret the expression $\Tr(\rho X)$ as the mean value of the observable $X$. This expression is nothing else than the expectation value of a quantum-mechanical measurement of the observable $X$ on a system in the state $\rho$. Consequently, the probabilities required here are just quantum-mechanical probabilities\footnote{As \citet[p. 25]{Wallace2016} puts it: \ZT{there \textit{is} something probabilistic about $\rho$, and about the forward-compatibility requirement, but only in the sense that there is something probabilistic about the quantum state itself (however that probabilistic nature is to be understood)}.}, and if one assumes that the Everett interpretation can explain probabilities, then it is also able to explain the probabilities in statistical mechanics. In particular, if no measurements are made, there is no need for probabilities since then there is nothing probabilistic about the formalism introduced in section (\ref{mz}) - we are simply solving differential equations and making approximations for them. Strictly speaking, the question we should be asking when confronted with the formalism presented in section (\ref{mz}) is not \ZT{What is the meaning of the probability distribution in statistical mechanics?}, but simply \ZT{What is the meaning of the symbol $\rho$?} This can be answered with \ZT{the density operator} regardless of whether or not we take the density operator to represent probabilities. (Note that the we do not need to adapt the Everett interpretation here. \textit{Any} interpretation of quantum mechanics that allows to interpret $\Tr(\rho X)$ as the expectation value of a measurement of $X$ on a system in state $\rho$ - in other words: any interpretation in which the Born rule holds - allows for such an understanding of $\rho$. Just pick whatever is your favourite interpretation of quantum mechanics.)

Second, one could ask what this implies for classical statistical mechanics, which the MZ formalism is also applied to. While in a real physical system one could argue that it is always ultimately described by quantum mechanics\footnote{\citet{Wallace2016} argues, in particular, that the classical phase-space distribution function arises as a limit of the Wigner function, which is equivalent to the density operator.}, thermodynamic irreversibility is also observed in classical molecular dynamics simulations \citep{Toth2020} that involve no quantum effects of any sort. Moreover, one can apply the MZ formalism (like statistical mechanics in general) also to astrophysical \citep{teVrugtHW2021} or colloidal \citep{EspanolL2009} systems, and it is not very plausible that the dynamics of macroscopic colloids or even stars depends on quantum effects. Finally, one could argue that, if $\vec{\Gamma}_0$ is the actual microscopic value of the phase-space variables contained in a vector $\vec{\Gamma}$, the microscopic density is simply proportional to $\delta(\vec{\Gamma}-\vec{\Gamma}_0)$ with the Dirac delta distribution $\delta$. This is actually common practice in classical many-body physics \citep{teVrugtW2020b}. However, a density given by a Delta distribution will typically not take the form (\ref{relevantdensity}), and the assumption that the initial density has this form was quite essential for the derivation of Eq.\ (\ref{markov}).

To understand this issue, we should clarify what we mean by the mean value $a_i$ given by Eq.\ (\ref{average}). It is an ensemble average, and the approximate transport equation (\ref{markov}) describes the dynamics of the ensemble average of the observable $A_i$. The ensemble average, and the ensemble average only, is monotonously approaching equilibrium. In contrast, the actual value in an individual classical system will typically approach a state corresponding to the macrostate with the largest phase-space volume (Boltzmannian equilibrium), but will continue to fluctuate around this equilibrium state. A good way to see this is to consider the example of \textit{dynamical density functional theory} (DDFT) \citep{teVrugtLW2020}, a theory for the one-body density\footnote{By \ZT{one-body density}, I mean the number of particles at a certain position, not the microscopic probability density discussed in the rest of this article.} of a classical fluid that exists in deterministic and stochastic forms. The deterministic theory, which can be derived as a special case of Eq.\ (\ref{markov}) \citep{EspanolL2009} describes the ensemble-average of the one-body density, its stochastic counterpart describes actual physical systems \citep{ArcherR2004,teVrugtLW2020,teVrugt2020}. In deterministic DDFT, equilibrium is approached monotonously, whereas there are fluctuations around equilibrium in stochastic DDFT. If we take the microscopic distribution $\rho$ to be proportional to $\delta(\vec{\Gamma}-\vec{\Gamma}_0)$ in the classical case (such that the ensemble average of an observable is always just its actual value in one specific system), then $f_i$ will never vanish and the dynamics will never, strictly speaking, be forward predictable by the coarse-grained dynamics (which in practice typically implies that it fluctuates away from equilibrium). We can, of course, introduce a \ZT{smoother} microscopic distribution by hand, for example by considering a probability distribution over initial conditions (this is what is typically done in classical statistical mechanics). In a computer experiment, one can repeat a classical simulation several times with random initial conditions and calculate the average of an observable over all these simulations. However, the probability distribution this average is taken with respect to might has quite a different interpretation than the quantum density operator $\rho$, since it is not a physical property of an actual system and can therefore not be used to explain an actual system's behavior.

Although Wallace's position is very reasonable, it is not free from additional assumptions. To make them explicit, it is helpful to introduce (following \citet{Maudlin1998}, see also \citet{Jaeger2014,Naeger2020}) the distinction between the \textit{ray view} and the \textit{statistical operator view}. These are two different ways of representing the state of a quantum system. According to the ray view, the state of a quantum object is described by ray in Hilbert space, while according to the statistical operator view, it corresponds to a density operator. A good way to illustrate this difference is to use a singlet state of a two-particle system. In this state, it is not possible to attribute a definite Hilbert space ray to a single particle. On the ray view, this would imply that each of the individual quantum objects has no state at all (no-state mode) or that it has a state only relative to the other particle (relative-state mode). On the statistical operator view, on the other hand, one can ascribe a state also to each of the individual particles, namely by tracing the density operator of the complete system over the degrees of freedom of the other one. Note that the statistical operator view still implies a failure of reductionism since the density operator of the joint system cannot be reconstructed from those of the individual particles.

Wallace's position, which is based on resolving the mystery of probabilities by reducing them to the quantum state of the system under consideration, works only if the statistical operator view is correct. If the ray view would hold, then \ZT{the quantum state of the system} would in general not exist (more precisely, it would only exist if it corresponds to a pure state), and if it does not exist, it cannot serve as an explanation for the probabilities of statistical mechanics. In particular, \citet{Maudlin1998} argues that an isolated particle will, according to the ray view, have to be in a state corresponding to a Hilbert space ray, while philosophers of physics formulating the past hypothesis as a condition for the density operator \citep{Wallace2011,Chen2020} typically allow the universe to be in a mixed state. We have thus uncovered a hidden assumption in Wallace's interpretation of statistical mechanics. (\citet[p. 33]{Robertson2021}, whose approach to statistical probabilities is similar to that of Wallace, explicitly states that \ZT{the individual state of the system in QM is not represented by a ray in Hilbert space (the quantum equivalent of a point in phase space), but a density matrix}, making the commitment of this approach to the statistical operator view even more apparent.)

However, I would argue that this result is not detrimental to Wallace's view. Instead, if we take into account how well it resonates with the formalism employed in modern physics, and if we adopt the plausible principle that we should (if in doubt) always adopt the interpretation that fits best to the practice of physicists, then it would follow that Wallace's interpretation actually provides an argument in favor of the statistical operator view. Moreover, as emphasized by \citet{Chen2020} (who develops, based on a density operator realism, an understanding of the past hypothesis as a specific assumption about the initial density operator of the universe), we can do quantum mechanics (including variants such as Bohmian mechanics) just as well based on the density operator as we can based on the wavefunction. Consequently, I will adopt here both Wallace's interpretation of $\rho$ and the statistical operator view.

However, this still leaves us with the question what $\bar{\rho}$ is. In the presentation by \citet{Wallace2016}, $\bar{\rho}$ is simply what one gets if one applies the projection operator $P$ to $\rho$. Formally, this is absolutely correct. However, it does not mention an important point about why one constructs the relevant density and the projection operator in the way one does. To see why one has to coarse-grain in this particular way, we have to consider why Eq.\ (\ref{relevantdensity}) has the form it has. The reason is, as discussed in section (\ref{mz}), information theory. The relevant density is, in the spirit of Jaynes, constructed by maximizing the informational entropy. Consequently, while $\rho$ is an ontic probability (or better: state), $\bar{\rho}$ is an epistemic probability distribution even in the quantum case. The question whether probabilities in statistical mechanics are epistemic or ontic thus has a surprising answer: both.

Thus, we have found an account of probability of that combines the first and second interpretation suggested in section (\ref{probs}), by taking $\rho$ to be an ontic (quantum-mechanical) and $\bar{\rho}$ to be an epistemic (information-theoretical) probability. This account will be referred to as \textit{option (a)}. An alternative account - which will be referred to as \textit{option (b)} - can be constructed based on the third interpretation from section (\ref{probs}), namely Myrvold's theory of almost-objective probabilities. Here, the initial probability distribution represents one's initial credences about the system. The credences at later times will not be the Hamiltonian time evolutes of the initial one, but instead will be simpler distributions determined by the macrostate of the system. Thus, Myrvold's theory also involves - for a given observer - two probability distributions, namely the Hamiltonian time evolute of the initial credences and the simpler distribution used at later times. We may identify these two distributions with the two distributions appearing in the MZ formalism by assuming $\rho(t)$ to be the Hamiltonian evolute of the initial credences at time $t$ - in line with the fact that, in section (\ref{mz}), we have assumed that $\rho$ evolves according to Eq.\ (\ref{liouville}) - and $\bar{\rho}(t)$ with the actual credences our observer has at time $t$. In particular, this forces us to set $\rho(s)=\bar{\rho}(s)$ (since our observer has only one credence function at the initial time). Next, we observe that $\rho$ and $\bar{\rho}$ evolve differently. In Eq.\ (\ref{exact}), the organized drift term $v_i$ is the part of the dynamics that we would have if $\rho$ evolved like $\bar{\rho}$ at all subsequent times, whereas the memory terms (that lead to dissipation and thus equilibration) result from deviations of $\rho$ and $\bar{\rho}$ \citep{Grabert1978}. Consequently, in a Myrvoldian framework, we can indeed interpret $\rho(t)$ as the time evolute of our initial credences, and $\bar{\rho}$ as the simpler distribution that we use as a surrogate for $\rho$. This interpretation is very different from the one suggested in \citet{Grabert1982}, where $\rho$ is the observer-independent microscopic distribution. In fact, it would here be assumed that $\bar{\rho}$ (and not $\rho$) is \ZT{quasi-objective} in the sense defined in section (\ref{probs}). However, this interpretation is also compatible with the derivation presented in section (\ref{mz}). 

I wish to emphasize here that option (b) is not Myrvold's theory, but my own proposal for how Myrvold's theory might be combined with the formalism presented in section (\ref{mz}). One might object here against the view that we have to set $\rho(s)=\bar{\rho}(s)$ on this account. After all, Myrvold's approach is based on the method of \textit{arbitrary} functions, and it is quite essential that any initial credence function will lead to a similar stationary (equilibrium) state as long as it is reasonable. \ZT{Reasonable} here means that if can arise from macroscopic measurements and does not require us to postulate detailed knowledge about microscopic correlations like those required for generating anti-thermodynamic time evolutions. Surely this initial credence does not have to be given by Eq.\ (\ref{relevantdensity}). However, although (\ref{relevantdensity}) is by far the most common choice, one could construct $\bar{\rho}$ in a different way as long as it satisfies macroequivalence and can be written as a function of rhe macroscopic variables only. In fact, \citet[p. 20]{Grabert1982} even argues that the condition $\rho(s)=\bar{\rho}(s)$ \ZT{should be looked upon as a condition for an adequate definition of the relevant probability density rather than a restriction on initial states}. An \ZT{unreasonable} density can presumably not be written as a function of the relevant variables. For all others, we may indeed construct $\bar{\rho}$ in such a way that it matches our initial credence function at time $s$. (In this work, I will nevertheless - as is common and done for most of Grabert's book - always use the form (\ref{relevantdensity}).) 

The problem with option (b) is that it is assumed in the MZ formalism that, for \textit{any} observable $X$, the expectation value is given by $\Tr(\rho X)$ - not just for the macroscopic observables, for which $\bar{\rho}$ gives the correct expectation value. This is not guaranteed if $X$ is in any way related to the credences of a human observer. In contrast, in quantum mechanics, one can simply take $\rho$ to be the density operator of the system of interest, which in general will be a mixed state. Moreover, only option (a) allows us to use the initial form of $\rho$ as an explanation for the physical behavior of a system (this issue will be discussed further in section (\ref{fifthproblem}). For these reasons (and because it is closer to what is actually assumed in physics), I will adopt option (a) in this work. When appropriate, I will also discuss possible implications that choosing option (b) instead would have for the discussion here.

\section{\label{secondproblem}Second problem: Definition of equilibrium and entropy}
Now, we are in the position to address the problem(s) of irreversibility discussed in (\ref{coarse}). Since, the first one is mainly concerned with phenomenological thermodynamics, I will start with the second one, the question of how to define equilibrium and entropy.

In a closed Hamiltonian system, the density operator $\rho$ will never approach the equilibrium form if it did not start in equilibrium. Broadly speaking, there are (at least) three ways around this issue. The first one (not considered here, but in a certain variant refuted in \citet{teVrugtTW2021}) is to argue that the microscopic dynamics of a closed system is not actually Hamiltonian. The second one is that external perturbations destroy correlations and thereby lead to an equilibrium form for $\rho$ (\textit{interventionism}, see \citet{RidderbosR1998}). The third one (adopted for most of this work) is to define equilibrium based on the coarse-grained density $\bar{\rho}$. This relevant density does indeed approach an equilbrium form for the Markovian dynamics (\ref{markov}), which is shown for specific application scenarios in \citet{EspanolL2009} and \citet{AneroET2013}. Thus, we may characterize an isolated system as being in equilibrium if $\bar{\rho}$ has an equilibrium form. Note that the second and third option differ solely in the way they define \ZT{equilibrium}, but not with regards to the observable processes they predict for any physical setup \citep{teVrugt2020}. Moreover, a definition of \ZT{equilibrium} in terms of $\bar{\rho}$ can also capture an interventionist understanding, since we may take $\bar{\rho}$ to describe the degrees of freedom of a subsystem that is coupled to an external heat bath \citep[p. 72]{Grabert1982}.

Let us thus follow the idea that a system is in equilibrium if $\bar{\rho}$ takes the equilibrium form, which would essentially correspond to Jaynes' view. A common objection against this position (see \citet{Frigg2008}) is that, if $\bar{\rho}$ is a representation of our knowledge rather than of the system's state itself, defining \ZT{equilibrium} in terms of $\bar{\rho}$ would imply that it is our knowledge that is in equilibrium rather than the system itself. Similarly, it would be our knowledge that approaches equilibrium and not the system itself. 

Three responses can be made. The first one, coming from Jaynes himself, is that \ZT{entropy} is indeed an epistemic concept. It is not directly measurable, and which variables it is a function of depends on which variables the experimenter has chosen to be relevant \citep{Frigg2008}. Evidently, this is also the case for the entropy (\ref{entropy}). This view is plausible by the Gibbs paradox, which shows that whether a mixing entropy is to be assigned to a certain mixing process depends on whether we decide to view it as a change of the system's state (see the brief discussion in \citet{teVrugt2021b}).

Second, $\bar{\rho}$ (and thereby an entropy/equilibrium defined in terms of it) is not \textit{solely} something that depends on our knowledge. It depends on the mean values $\{a_i\}$, which (at least if we take a \ZT{mean value} to be a quantum-mechanical expectation value) are objective physical features of the system. In nonequilibrium, these expectation values depend on time, whereas in equilibrium, they reach a stationary state. Once I have chosen a set of relevant variables, the statement \ZT{all relevant variables are constant} is a physical statement that can be checked experimentally. Similarly, as discussed in Section (\ref{fourthproblem}), it is a nontrivial physical statement that the H-theorem (\ref{htheorem}) holds. All of this has to do with the fact that there is still the objective density operator\footnote{I write \ZT{objective density operator} to emphasize that I take the density operator to be a feature of the physical world rather than a representation of our ignorance about the actual pure state.} $\rho$ in the background that we aim to make inferences about.

While these two responses are based on option (a), a third one could be given when using option (b) (which is based on Myrvold's idea of almost objective probabilities). \citet[p. 594]{Myrvold2016} proposes to use the microcanonical distribution \ZT{as a surrogate for the evolute of our initial probability distribution} as it gives correct results for all feasible measurements. On this account, a system is in equilibrium if an equilibrium distribution gives correct predictions for all feasible measurements. When using option (b), $\rho(t)$ - which will not have a microcanonical form - is the time evolute of the initial probability distribution, and $\bar{\rho}(t)$ is what we use instead for convenience. Consequently, if $\bar{\rho}$ takes an equilibrium form after a sufficiently long time, we have indeed reached equilibrium.

\section{\label{thirdproblem}Third problem: Coarse-graining}
The next issue is the justification of coarse-graining. \citet[p. 561]{Robertson2018} has proposed that this issue can be split into two sub-problems - namely, the justification of a \textit{particular} coarse-graining projection and the justification of coarse-graining \textit{in general}.

As discussed in section (\ref{mzp}), \citet{Robertson2018}, takes the revelation of autonomous macro-dynamics to be the justification for coarse-graining in general. A particular coarse-graining method then should be chosen in such a way that the system obeys an autonomous dynamics on the macrolevel. She explicitly admits that \ZT{this criterion will not help physicists discover new, useful maps}, and that the resulting projection operators \ZT{will not look especially unified} \citep[p. 568]{Robertson2018}. This, however, is way too pessimistic, since (as shown in section (\ref{mz})) the projection operator \textit{can} be specified in a unified way, namely by Eq.\ (\ref{projectionoperator}). Similarly, the relevant density \textit{does} have a unified form, namely the one given by Eq.\ (\ref{relevantdensity}). This is an interesting observation since this form is constructed based on Jaynes' information-theoretic approach. 

The question is then what counts as \ZT{autonomous macro-dynamics}. By \ZT{autonomous dynamics}, \citet[p. 553]{Robertson2018} means that the dynamics of $\bar{\rho}$ depends neither on $\delta\rho$ nor (explicitly) on $t$ . The explicit time dependence is eliminated by the Markovian approximation, the $\delta\rho$ dependence by the assumption $\delta\rho(s)=0$. This definition of \ZT{autonomous}, which is the one used in the theory of dynamical systems, is somewhat unfortunate in this context as it would (combined with the idea that the MZ formalism ought to reveal autonomous dynamics) automatically render all applications of the MZ formalism to systems driven by explicitly time-dependent external potentials \citep{teVrugtW2019} unjustified. Presumably, however, we can understand Robertson's criterion as implying that we should choose the $\{A_i\}$ in such a way that their mean values obey an equation of the form (\ref{markov}), i.e., that the Markovian approximation is justified. This requires that the relevant variables are slow compared to the microscopic degrees of freedom. A set of slow variables can typically be constructed by considering both the conserved variables and the variables associated with a spontaneously broken symmetry. As an example, consider the dynamics of a crystal. Here, the slow variables are density and momentum (conserved variables), and the symmetry-restoring low-frequency Goldstone modes of the crystal. Consequently, these are an appropriate set of relevant variables for deriving a theory for the elastic properties of a crystal via the MZ formalism \citep{WalzF2010,RasSF2020}. This shows that Robertson's criterion is not only (in contrast to what is claimed in \citet{Robertson2018}) useful for practitioners of physics, it is actually already used by them.\footnote{This statement should be considered a compliment to rather than a criticism of Robertson's results. Providing a conceptually clean analysis of the practice in physics is a meritorious achievement for a philosopher of physics.} 

Moreover, Robertson argues that revealing autonomous macrodynamics is \textit{the} justification for coarse-graining, while justifications like measurement imprecision are inappropriate. The reason is, \citet{Robertson2018} argues, one would otherwise face the problem that irreversibility (an effect associated with coarse-graining) would be illusory and/or anthropocentric. As she puts it \citep[p. 565]{Robertson2018}, it \ZT{seems unlikely that advances in the science of microscopy will lead to different choices of} (the projection operator). This view, again, is related to the fact that philosophers tend to study coarse-graining almost exclusively in its relation to irreversibility, thereby ignoring its much wider use in physics. 

A good example for an application of the MZ formalism that is not related to autonomous macrodynamics but to the limitations of human observers is the study of turbulent fluids. Their dynamics is characterized by a coupling of all length scales (i.e., all wavenumbers) in the system. Small length scales influence the large ones and vice versa. When studying and simulating such a fluid - a problem of great importance in engineering - one faces the problem that simulations can only resolve a finite length scale. This finite resolution then leads to inaccuracies in the simulation results also on large scales.

This problem is frequently addressed using so-called \ZT{large eddy simulations} (see \citet{Sagaut2006} for an introduction). In a large eddy simulation, one simulates the large scales explicitly and includes the smaller scales via a subgrid model. Here, the MZ formalism can play a useful role \citep{ParishD2017,MaeyamaW2020}. One uses the Navier-Stokes equation (which describes the dynamics of incompressible fluids) as a microscopic model and then projects onto the small wavenumbers (i.e., the large length scales), which are the relevant variables. The memory terms then incorporate the small-scale effects. 

Notably, this is not done because the large length scales in the fluid obey an autonomous macrodynamics. In fact, it is precisely the problem that they do not. Instead, we use the MZ formalism because our computers are not good enough to solve the full Navier-Stokes equation numerically, i.e., because of limited available microscopic information. Of course, this coarse-graining induces artefacts, which can be seen as anthropocentric. This, however, is unavoidable (although one of course wishes to minimize it). And if \ZT{better microscopes} (i.e., better computers) would be available, windpark engineers would certainly use them rather than the less accurate LES models.

Similar ideas are relevant for general relativity. Since the Einstein field equations are highly nonlinear and therefore do not commute with an averaging procedure, it is not possible to get an averaged large-scale model of the universe by simply inserting the averaged matter distribution into these equations. However, this is precisely what is done in the derivation of the Friedmann equations. This issue, known as the \ZT{averaging problem}, is not fully understood and has even been suggested as an explanation for dark energy \citep{ClarksonELU2011}. Recently, \citet{teVrugtHW2021} have addressed this problem by extending the MZ formalism to general relativity, which allowed them to derive a correction term for the Friedmann equations. As in the case of turbulence, this study is motivated not by the existence of autonomous macrodynamics but by the impossibility of actually solving the Einstein field equations for the complicated matter distribution of the universe.

This does not mean that Robertson's justifying of coarse-graining in the case of irreversible transport equations in nonequilibrium statistical mechanics is not correct - it is fully appropriate for the analysis of irreversibility. The point I wish to make here is that the justification of coarse-graining depends heavily on the context in which it is used, and that \ZT{measurement imprecision} is not an illegitimate one - it is necessary to use procedures of this form for designing airplanes or windparks. Studying the justification of coarse-graining case by case is important as it has implications for our understanding of the effects that result from it. For example, the predictions that MZ-based models with simple approximations for the memory kernel make for transfer spectrum in turbulent flows can differ from the actual spectrum \citep[p. 17]{ParishD2017}. This is, like irreversibility, an effect that arises only after coarse-graining. However, in the case of irreversibility, we can - as shown by \citet[pp. 573-576]{Robertson2018} - consider it to be not illusory, but (weakly) emergent. It is a consequence of robust autonomous macrodynamics. In contrast, the artefacts in the MZ-based large eddy simulations are not emergent, but simply an (unavoidable) technical error. This is due to the fact that we coarse-grain here not to reveal autonomous macrodynamics, but simply because our human and computational limitations leave us with no other option.

To summarize: We coarse-grain because we wish to study the subdynamics of a certain set of variables in a system we cannot (or do not want to) describe completely. This can be done because of human or technical limitations - as in the case of turbulence - or because we wish to reveal or study autonomous macro-dynamics - as in the case of irreversible statistical mechanics.

\section{\label{fourthproblem}Fourth problem: Approach to equilibrium}


I have thus argued that coarse-graining in statistical mechanics has an information-theoretic basis. It is based on what we know about the system or what we are interested in, and it changes if we know more or if we are interested in more. This raises a question: \citet{Robertson2018} introduces \ZT{the possibility of revealing autonomous macro-dynamics} as the justification for a particular form of coarse-graining to avoid the problem that, if irreversible equations of motion are found by ignorance-based coarse-graining, irreversibility might be an illusion. Given that I now propose that coarse-graining is based on information theory and that $\bar{\rho}$ represents an epistemic probability distribution, does that imply that I have precisely this problem? The answer is no, and the reason is that Robertson, while not fully answering the third question, is correct about the fourth one. While the set of relevant variables $\{A_i\}$ can in principle be chosen in an arbitrary way, not every such set will be found to obey a closed macroscopic dynamics. 

It is worth briefly recalling here in which way the memory term generates (in the Markovian limit) irreversible dynamics (following \citet[pp. 62-65]{Zeh1989}). The relevant information present initially is, by the operators appearing in the memory kernel, transformed into irrelevant information. This initially formed irrelevant information corresponds to so-called \ZT{doorway states} (for example two-particle correlations). The subsequent application of the propagator $G(s,t)$ transforms this irrelevant information deeper into the irrelevant channel (for example by creating many-particle correlations). Due to the depth of the irrelevant channel (in a system with a large number of particles), it takes an almost infinitely long time (recurrence time) for the information to come out of the irrelevant channel again, ensuring that the \ZT{irrelevant information} is indeed irrelevant for the dynamics of the relevant degrees of freedom. The Markovian approximation in particular assumes that the relevant variables evolve so slowly that they do not change during the time it takes for the irrelevant information to move from the doorway states into the irrelevant channel, ensuring that one can effectively assume that there never is irrelevant information and that we can therefore write a closed dynamics for the relevant degrees of freedom.

The Markovian approximation is the step that shows most clearly why the irreversibile equations obtained via the MZ formalism are objective. It corresponds to the assumption that the macroscopic variables evolve slowly compared to the microscopic ones. This is is a \textit{physical} assumption about what the system actually does, and whether we observe irreversibility depends on whether this assumption is true. The importance of the Markovian assumption can be seen by considering examples in which this assumption is \textit{not} satisfied.

An application of the MZ formalism that is very important in physics but essentially ignored in philosophy of physics is the derivation of mode coupling theory (MCT). This method is used to model the behavior of glassy systems \citep{Das2004}. Roughly speaking, glasses form when particles in a dense undercooled liquid get trapped such that they cannot move to their equilibrium positions in a crystal (\ZT{caging}). This prevents the system from reaching its equilibrium state (which would correpond to a crystal), leaving it in a disordered state with strong dependence on the history of the system (\ZT{aging}) instead.

In the derivation of MCT, one projects onto density and current, which are typical slow variables for a fluid \citep{Janssen2018}. This allows to derive a formally exact equation of motion for the density correlator $\phi_q$ involving memory effects. Instead of just dropping them completely, one makes a simple ansatz for the form of the memory kernel by expressing it via the time correlation of products of density modes. It might be surprising that products of density modes are among the irrelevant variables given that we have chosen the density as a relevant one. This is a consequence of the fact that, in the Hilbert space of dynamical variables, $A$ corresponds to a different direction than $A^2$ (where $A$ is an observable) \citep[p. 151]{Zwanzig2001}. Nevertheless, if $A$ is slow, it is of course not unlikely that $A^2$ is also slow. This is precisely what happens in MCT \citep[p. 6]{Kawasaki2009}: Since the irrelevant variables (which include quadratic density fluctuations) are also slow, they cannot be ignored, such that the final equation of motion also contains memory. (In Zeh's terminology: the system remains in a doorway state.) For small couplings, $\phi_q$ goes to zero for $t\to\infty$, which means that an initial density perturbation vanishes after a sufficiently long time. However, for larger coupling constants, $\phi_q$ remains finite at all times, and the liquid does not go to equilibrium \citep[p. 878]{Goetze1998}. The system has undergone a transition to a nonergodic state \citep{FuchizakiK2002}.

The fact that a system described by simple MCT might never reach equilibrium despite the fact that it is described using the same information-theoretic coarse-graining methods that are used in other systems shows that coarse-graining by itself does not lead to dynamics that exhibit an approach to equilibrium. What is relevant instead is that the memory kernel decays quickly such that a Markovian approximation is possible. Hence, at least as long as we discuss applications of the MZ-formalism to reversible Hamiltonian dynamics, the fourth problem essentially corresponds to the question why a Markovian approximation is possible. 

This, it turns out, is a very interesting and not fully solved physical problem. Recalling the three options introduced in section (\ref{secondproblem}), we can consider here (1) non-Hamiltonian microdynamics, (2) interventionism, and (3) coarse-graining. A coarse-grainer who makes the Markovian approximation when describing a closed system will understand this approximation as, strictly speaking, describing \ZT{pseudo-irreversibility} since the recurrence time (the time after which a Hamiltonian system returns (arbitrarily close) to its initial state), which becomes extremely large for macroscopic systems, is assumed to be infinite \citep{Toth2020}. Markovian transport equations of the form (\ref{markov}) do not only predict irreversibility, they also make a specific quantitative prediction for the way equilibrium is approached - in hydrodynamic systems typically in a diffusive manner (as is confirmed by experiment). In a recent study, \citet{Toth2020} has investigated via computer simulations whether pseudo-irreversibility is present in a closed many-body system. While the answer turned out to be yes, the precise relaxation behavior was not diffusive. As possible reasons, \citet{Toth2020} discusses the absence of finite-scale terms and a possible non-Hamiltonian microdynamics. The problem of (non-)diffusive relaxation provides an interesting problem also for the philosophy of statistical mechanics since its resolution might lie in one of the three options mentioned above - diffusive relaxation could (1) result from a (yet to be discovered) deviation of the actual microdynamics from a Hamiltonian form, it could (2) actually only exist in non-isolated systems, and (3) it might arise from Hamiltonian microscopic dynamics in a way that is yet to be understood more precisely. This, I would argue, is the debate that interventionists and coarse-grainers should actually be having since whether diffusive relaxation exists in a closed system is something one can make different predictions for (whereas the actual discussion they are having, namely whether there can be equilibration in a closed system, is, as mentioned in section (\ref{secondproblem}), a matter of terminology - both interventionists and coarse-grainers will accept that isolated systems typically approach a homogeneous stationary state, the question is solely whether we should call this state \ZT{equilibrium}). Once again, we can see that philosophers of physics should pay closer attention to quantitative theories for the approach to equilibrium.

Note that the connection between irreversibility and the Markovian approximation changes if the microdynamics is not Hamiltonian. For example, if we apply the MZ formalism to a system with dissipative microscopic dynamics (an example would be the derivation of Green-Kubo relations for chiral active matter by \citet{HanFSVIdPV2020}), then the organized drift $v_i$, which is always time-reversal invariant in the Hamiltonian case, can already describe equilibration. Usually, this occurs because one applies the MZ formalism to a \ZT{microdynamics} that is already coarse-grained, such as the Langevin equations describing the motion of colloidal particles in a fluid providing friction. 

Finally, let us consider how this problem looks from the perspective of option (b). In this case, the Markovian approximation is just the mathematical representation of what \citet{Myrvold2012} suggests the explanation of equilibration in his approach to be, namely (in our terminology) the mechanisms that make the difference between $\rho(t)$ and $\bar{\rho}(t)$ irrelevant at late times. The Markovian approximation corresponds (in the variant of the MZ formalism considered here) to the assumption that the system quickly relaxes to the constrained equilibrium given by (\ref{relevantdensity}). (Interestingly, this assumption tends to lead to an underestimation of relaxation times compared to simulations \citep{Kawasaki2006b}). At the same time, the Markovian approximation is what leads to irreversibility. Consequently, the mechanism that brings one to equilibrium and the mechanism that makes differences between $\rho$ and $\bar{\rho}$ irrelevant is indeed the same, and it is given by the slow time evolution of the relevant variables. Notably, this slow time evolution is a physical fact, such that the approach to equilibrium is a (quasi-)objective phenomenon also in this account. (In a later article, \citet{Myrvold2016b} provides a brief discussion of the MZ formalism and explicitly links the irreversibility produced there to local equilibration.)




\section{\label{fifthproblem}Fifth problem: Arrow of time}
We now turn to the final problem, namely the arrow of time. The discussions in \citet{Wallace2011} and \citet{Robertson2018} suggest to analyze this problem based on the idea of \ZT{forward compatibility} (see section (\ref{mzp}), a condition that is satisfied if the microscopic and the $C+$ dynamics agree, and that (according to the simple dynamical conjecture) is satisfied for simple initial densities. These then explain the arrow of time. Consequently, I will start this section with a mathematical analysis of the idea of forward compatiblility within the framework of the MZ formalism.

Recall that the $C+$ dynamics works by starting from an initial density $\rho(s)$, projecting it onto $\bar{\rho}(s)$, evolving it forwards in time for a small time interval $\Delta t$ using the microdynamics $U$, applying the projection $P^\dagger$ again, evolving it forwards again and so on. Due to Eq.\ (\ref{property}), we can then assume at each time $t$ that the density at time $t-\Delta t$ was of the relevant form (since at this time we have applied the projection operator to eliminate every other part of the density). As shown in section (\ref{mz}), the assumption that $\rho(s)=\bar{\rho}(s)$ allows to set $f_i(t,s)=0$. Let us use the fact that $\rho(t-\Delta)=\bar{\rho}(t-\Delta t)$ if $\rho$ is evolved via the $C+$ dynamics. Then, Eq.\ (\ref{exact}) gives (setting $s=t-\Delta t$)
\begin{equation}
\dot{a}_i(t)=v_i(t) + \INT{t-\Delta t}{t}{u}R_{ij}(t,u)a_j^\natural(u). 
\label{cplus}
\end{equation}
Hence, in the $C+$ dynamics, we can calculate $\dot{a}_i(t)$ by using an extremely short memory kernel (the memory integral only covers a time $\Delta t$). There are two assumptions we have to make in order for the $C+$ result (\ref{cplus}) to agree with the exact dynamics given by Eq.\ (\ref{exact}):
\begin{enumerate}
    \item The memory kernel has to fall off on a very short timescale (namely $\Delta t$), such that it does not matter that we have eliminated most of the memory integral.
    \item The mean random force $f_i$ has to vanish.
\end{enumerate}
If we compare these two assumptions to the two approximations we have made in section (\ref{mz}) to arrive at the irreversible dynamic equation (\ref{markov}), we can see that they are exactly the same. A rapidly decaying memory kernel implies Markovian dynamics, and a vanishing mean random force is the result of $\delta\rho(s)=0$. This result teaches us two important lessons:
\begin{enumerate}
    \item The simplicity of the initial density ($\rho(s)=\bar{\rho}(s)$) is required for forwards-compatibility, as suggested by the simple dynamical conjecture.
    \item A simple initial state is not sufficient for forward compatibility, as it does not by itself allow for a Markovian approximation. In addition to a condition on the initial state (simplicity) to solve the fifth problem, we also require a condition on the dynamics (quickly relaxing memory kernel) to solve the fourth problem.
\end{enumerate}
In principle, this result should not be surprising. For a Hamiltonian system where the recurrence time is very short, there is obviously no way to get irreversible dynamics by just imposing the \ZT{right} initial condition. (\citet[p. 292]{Wallace2015}, in his discussion of the MZ formalism, also notes that one requires both a time-symmetric constraint on the dynamics and a constraint on the initial state.) Since the Markovianity condition has already been discussed in section (\ref{fourthproblem}), we can now turn to the other one ($\rho(s)=\bar{\rho}(s)$), which fixes the (thermodynamic) arrow of time.

Conceptually (based on the idea of \ZT{irrelevant information channels} discussed in \citet{Zeh1989} and briefly reviewed in section (\ref{fourthproblem})), the symmetry breaking provided by the assumption $\rho(s)=\bar{\rho}(s)$ can be understood as follows: At the initial time, there is no irrelevant information ($\rho=\bar{\rho}$), and irrelevant information generated subsequently goes into the irrelevant channel and therefore does not affect the relevant dynamics. If we time-reverse this process, then the irrelevant information would come back out of the irrelevant channel and become relevant (and thus affect the time evolution of the macroscopic variables). Therefore, the irrelevant information is irrelevant only for predictions, but not for retrodictions. Moreover, the time evolute of a simple distribution is not simple since irrelevant information is created from the relevant one. Evolving the system backwards from the time $s$ at which we imposed simplicity, the relevant information (all that there is at time $s$) is also transferred into the irrelevant channel. Suppose now that we had imposed the condition $\rho=\bar{\rho}$ at the end rather than at the beginning of the process that we wish to study. Then, irrelevant information has to be present prior to the end and has to transform into relevant information during the time evolution (and therefore has to affect the macroscopic dynamics). Consequently, the macroscopic dynamics is, in this case, affected by microscopic many-particle correlations. This is precisely what happens both in simulations where anti-thermodynamic behavior is observed (such as the ones by \citet{teVrugtTW2021}, who artificially generated a highly correlated initial state for this purpose) and in real spin systems \citep{MicadeiEtAl2019} where anti-thermodynamic behavior arises due to initial correlations relevant for the subsequent time evolution.

Regarding the problem of symmetry breaking, \citet{Wallace2011} notes that a simple density is compatible not only with the forward-dynamics, but also with the backward dynamics induced by a given coarse-graining procedure. The problem is that, while the forward coarse-grained dynamics is usually accurate, the backward coarse-grained dynamics is not. Moreover, the forward time evolution of a simple distribution is not simple. Hence, simplicity can only be imposed once. He then discusses two choices for the time at which it is imposed, namely
\begin{enumerate}
    \item at the beginning of the process that one wishes to study.
    \item at the beginning of time.
\end{enumerate}
\citet[p. 21]{Wallace2011} relates the first option to Jaynes' objective Bayesian approach, while \citet[p. 559 - 560]{Robertson2018}, who discusses the same options, relates it to the practice of actual physics. Wallace then quickly dismisses option 1 based (among other things) on the argument that it would imply anti-thermodynamic behavior before the start of the process. Option 2, in contrast, ensures that problematic backward coarse-graining is not possible. Hence, this option should be chosen for explaining thermodynamic irreversibility. \citet[p. 560]{Robertson2018} notes that there will not be huge empirical differences between the predictions both options lead to.

While imposing simplicity at the beginning of time is indeed a reasonable way of explaining the universe's arrow of time, Wallace is in fact too quick with option 1. As discussed by \citet[p. 492]{Grabert1978}, the assumption that Eq.\ (\ref{relevantdensity}) gives the initial condition for $\rho$ is satisfied if the system starts in a state of constrained equilibrium, where (due to the application of external forces) the values of the macrovariables are forced to assume certain values. The microscopic degrees of freedom then relax towards the state that maximizes the system's entropy with respect to the macrosocopic constraint given by these macrovariables. At the start of the experiment (time $s$), the external field is removed, and the system starts to evolve from the simple distribution that was forced upon it as an initial condition. In this context (which is quite typical for simulations and experiments), it is very reasonable to impose simplicity at the beginning of the process we wish to study.

What about the objection that this predicts anti-thermodynamic behavior before the beginning of this process? This objection assumes that the microscopic dynamics is the same before this time. However, before the beginning of this process (i.e., during the preparation of the experiment), the system was subject to external forces, and these external forces modify the Hamiltonian. Let us, following \citet[p. 29]{Grabert1982}, assume that the system's own Hamiltonian is $H$ and that the external forces $h_i$ couple in such a way that they change the Hamiltonian to $H-h_i A_i$. Then, the system will relax to a generalized canonical state of the form
\begin{equation}
\rho = \frac{1}{Z}e^{-\beta(H- h_i A_i)}
\label{generalizedcanonical}
\end{equation}
with the rescaled inverse temperature $\beta$. The external forces are then switched off at the beginning of the process, and we observe how the system relaxes back to equilibrium. Evidently, (\ref{generalizedcanonical}) is a state of the form (\ref{relevantdensity})\footnote{We have written \eqref{generalizedcanonical} in a canonical form here, it can be transformed to the form (\ref{relevantdensity}) by including $H$ in the set of relevant variables \citep{Grabert1982}.}, and we are thus justified in assuming $\rho(s)=\bar{\rho}(s)$ (simple initial distribution). Nevertheless, this simple initial distribution arose precisely \textit{because} of normal thermodynamic behavior (relaxation to the state (\ref{generalizedcanonical}), which was the equilibrium state while the forces $h_i$ were still present).

A more sophisticated objection would run as follows: The reason we expect a state of constrained equilibrium for systems prepared in this way is that the microscopic degrees of freedom will relax to a maximum-entropy state subject to these external constraints. This, however, is already an irreversible process. Consequently, we cannot use this assumption to explain the arrow of time. However, this is not what most physicists intend to do (the constrained-equilibrium-assumption is even applied to systems where an external drive is switched on at the beginning of the experiment, which implies non-thermodynamic behavior \citep{teVrugtW2019,MenzelSHL2016}). While explaining the arrow of time presumably does require imposing simplicity at the beginning of time, it is perfectly reasonable to use option 1 if our goal is simply a quantitatively accurate description of a certain experiment. (Recall that, as \citet{Wallace2015} has noted himself, explaining the arrow of time is far from the only aim of nonequilibrium statistical mechanics.)

What we do have to note, however, is that assuming that $\rho$ is the objective density operator of the system under consideration - option (a) in the terminology introduced in section (\ref{probs}) - the assumption $\rho(s)=\bar{\rho}(s)$ implies that the initial constrained equilibrium is not coarse-grained, but fine-grained.\footnote{Note that Jaynes justifies the assumption $\rho=\bar{\rho}$ by arguing that we should choose the initial distribution in such a way that it maximizes the Shannon entropy subject to our knowledge, which is given by the macroscopic constraings (see \citet[pp. 255-258]{Sklar1995} and \citet{Frigg2008}). This quite reasonable from the perspective of option (b) from section (\ref{probs}), where $\rho(s)$ are our initial credences, but not helpful if we assume that $\rho$ is the objective density operator. Note, however, that Jaynes' approach should not be understood as explaining or trying to explain irreversibility \citep{Brown2017}.} The reason is that we make a \ZT{uniformity} assumption not (only) regarding $\bar{\rho}$, but regarding $\rho$ itself. Such an assumption can be justified by arguing that the system, while it is being prepared, is in contact with the environment and thereby subject to external perturbations. These then destroy correlations and ensure that the microscopic state assumes the form (\ref{relevantdensity}). This line of argument corresponds to being an interventionist (see section (\ref{secondproblem})) regarding the initial constrained equilibrium, but a coarse-grainer regarding the final equilibrium the system relaxes to. Since (as emphasized in section (\ref{secondproblem}) and in \citet{teVrugt2020}) the difference between interventionists and coarse-grainers is mostly a terminological one, such a mixture of views would not be unreasonable. 

If you do not like this mixture of coarse-graining and interventionism, you have two alternatives. First, you can read the assumption $\rho(s)=\bar{\rho}(s)$ in a more generous way as stating that the initial state $\rho$ is such that it does not contain \ZT{special} correlations that would lead to anti-thermodynamic behavior. In this case, the term $f_i$ in Eq.\ (\ref{exact}) should be and remain so small that it does not influence the macroscopic time evolution. Second, you can resort to option (b) (see section (\ref{probs})), in which such mixtures do not arise.

Finally, it is worth relating the past hypothesis as discussed here to the well-known problem of \ZT{disastrous retrodictions}. This problem is concerned with the fact that, if we apply the typical probabilistic reasoning of statistical mechanics to the past, we are led to the conclusion that entropy increases in both directions of time, in strong conflict with our actual records of the past. Consequently, we have to conditionalize on a past hypothesis (understood as the assumption that the early universe had a very low entropy) to avoid such a retrodiction \citep{Albert2000}. This way of justifying the past hypothesis has led many authors to the belief that we no longer need a past hypothesis if we just set up our probabilistic reasoning in such a way that it does not generate such retrodictions. For example, it was noted by \citet{Myrvold2016} that, while the microscopic Hamiltonian dynamics is time-reversal invariant, probabilistic reasoning of the type employed in statistical mechanics is typically not. In particular, we have memory of past, but no knowledge about the future. Myrvold argues that \ZT{Nothing can make our knowledge of the macrostate of the system at time $t_0$ irrelevant for \textit{retrodictions} about the state of the system at time $t_0$ or before}\footnote{The time $t_0$ in this quote is equivalent to the time $s$ in our notation.} \citep[p. 596]{Myrvold2016}, and that (in his approach) \ZT{there is no need to invoke a Past Hypothesis} \citep[p. 1]{Myrvold2012}. 

It is not incorrect to argue that probabilistic reasoning is (or at least can be) time-asymmetric. In fact, this is precisely the reason that, in Eq.\ (\ref{exact}), we integrate from the initial time $s$ forward in time to $t$, rather than backwards in time - the memory term is supposed to take into account knowledge about the previous states of the system \citep{Grabert1982}. However, it is certainly possible that a system exhibits anti-thermodynamic behavior (in fact, it can be observed in simulations using Hamiltonian microdynamics \citep{teVrugtTW2021}), and that such a behavior is never observed requires an explanation that appeals to the initial state of the system. Whether a system exhibits anti-thermodynamic behavior is a feature of this system quite independent of our probabilistic reasoning about it, and we cannot explain the absence of such behavior by any argument that is related to our probabilistic reasoning. If $\rho$ is a physical property of the system (option (a)), the question whether $\rho(s)-\bar{\rho}(s)$ (and whether we can thus set $f_i=0$ in Eq.\ (\ref{markov})) is equal to or different from zero has (once we have fixed $\bar{\rho}$) nothing to do with our probabilistic reasoning, it is just a physical boundary condition. Eliminating the need for such a boundary condition with arguments about probabilistic reasoning would be just as sensible as eliminating the need for a boundary condition in the explanation of the motion of a classical harmonic oscillator with arguments about probabilistic reasoning.

Note that, at this point, we have perhaps the most significant difference between options (a) and (b). As indicated in section (\ref{probs}), option (b) interprets $\rho$ as the time evolute of our initial credences, and therefore essentially forces us to set $\rho(s)=\bar{\rho}(s)$. This assumption is thus - unlike in option (a) - \textit{not} a physical boundary condition. Consequently, it cannot play an explanatory role in why the arrow of time has the direction it has. Moreover, $\bar{\rho}(t)$ is interpreted here as the convenient replacement we use for $\rho(t)$. There is little reason to use this replacement prior to time $s$, and so there are indeed no disastrous retrodictions when we use option (b). However, this absence of disastrous retrodictions is a disadvantage rather than an advantage. The reason is that there are systems that evolve away from equilibrium (such as correlated spins where heat flows spontaneously from a cold to a hot spin \citep{MicadeiEtAl2019}). For these systems, the result of a \ZT{disastrous retrodiction} would be correct. 

Similarly, \citet[p. 588]{Myrvold2016} has argued that applications of the principle of indifference to nonequilibrium systems generate disastrous retrodictions, since, starting from a distribution obtained in this way, entropy will increase in both directions of time. If this objection were correct, it would be a problem for MZ-based approaches using densities such as (\ref{relevantdensity}), which clearly are based on such a principle, in nonequilibrium. The reason why the MZ formalism does not have such a problem is that the principle of indifference is applied to $\bar{\rho}$, not to $\rho$ itself. The relevant density represents our information about the system and is based on the principle of indifference, whereas the actual density is a physical property of the system's (quantum) state. If we apply the principle of indifference to $\rho$ itself at a certain time $s$ (as we essentially do if we set $\rho=\bar{\rho}$ at the beginning of an experiment), then we do indeed have such a problem since we then would have an entropy minimum at time $s$ \citep[p. 67]{Zeh1989}. This can be avoided, e.g., by arguing that the system has interacted with the environment prior to $s$.


\section{\label{conclusion}Conclusion}
I have discussed in detail the derivation of time-asymmetric transport equations from time-symmetric microscopic dynamics in modern versions of the Mori-Zwanzig projection operator formalism. This has allowed for a qualitative and quantitative examination of various claims from the philosophical literature related to the status of probability and irreversibility in statistical mechanics that are based on \ZT{simpler} mathematical formalisms. In particular, I have shown that (a) the common dichotomy between epistemic and ontic approaches to probability in statistical mechanics is based on not distinguishing between the actual density operator $\rho$ and the relevant density $\bar{\rho}$, (b) that Myrvold's theory of almost-objective probabilities would correspond to interpreting $\rho(t)$ in a different way than in the physical literature (namely as the time evolute of one's initial credences), (c) that information-theoretical approaches are inevitable (and not harmful to the objective status of thermodynamics), (d) how Robertson's criterion of choosing a coarse-graining prescription corresponds to a rule for choosing the set of relevant variables $\{A_i\}$, (e) that, on the other hand, some sensible applications of the MZ formalism are \textit{not} based on this criterion, (f) that interventionists and coarse-grainers can find a new battle ground in the study of diffusive relaxation, (g) how Wallace's idea of forward compatibility can be accomodated within Grabert's MZ formalism, and (h) that quasi-objectivist interpretations of probability deviating from the ones used in physics avoid disastrous retrodictions (which is why they should not be used).

\begin{acknowledgements}
I am very grateful to Paul M. N\"ager for encouraging and supervising this project. I also wish to thank the participants of the thesis colloquium in theoretical philosophy - Ulrich Krohs, Oliver Robert Scholz, Peter Rohs, Ansgar Seide, and others - for helpful feedback on a previous version of this work. Moreover, I thank Raphael Wittkowski for introducing me to the MZ formalism, Michael Wilczek for interesting discussions about turbulence, Gyula I T\'{o}th for introducing me to the problem of diffusive relaxation, and Alice Rolf for assistance with the layout. Finally, I thank the Studienstiftung des deutschen Volkes for financial support.
\end{acknowledgements}

%
%


\end{document}